\documentstyle[multicol,prl,epsfig,aps,floats]{revtex}

\begin{document}
\addtolength{\topmargin}{2cm}
\bibliographystyle{prsty} 

\wideabs{
\title{Enhanced Stability of Layered Phases in Parallel Hard-Spherocylinders Due to Addition of Hard-Spheres}
\author{Zvonimir Dogic \and Daan Frenkel$^{\dag}$ \and Seth Fraden}
\address{The Complex Fluid Group, Martin Fisher School of Physics, Brandeis University, Waltham MA 02254
$^{\dag}$FOM-Institute for Atomic and Molecular Physics, Kruislaan
407, 1098 SJ Amsterdam, The Netherlands}
\date{\today}
\maketitle
\begin{abstract}
There is increasing evidence that entropy can induce microphase
separation in binary fluid mixtures interacting through hard
particle potentials. One such phase consists of alternating two
dimensional liquid-like layers of rods and spheres. We study the
transition from a uniform miscible state to this ordered state
using computer simulations and compare results to experiments and
theory. We conclude that (1) there is stable entropy driven
microphase separation in mixtures of parallel rods and spheres, (2)
adding spheres smaller then the rod length decreases the total
volume fraction needed for the formation of a layered phase,
therefore small spheres effectively stabilize the layered phase;
the opposite is true for large spheres and (3) the degree of this
stabilization increases with increasing rod length.
\end{abstract}
  }


\section{Introduction}

In hard particle fluids all allowed configurations have the same
energy and therefore it is the number of
 states, or equivalently the entropy of a system that determines
the equilibrium phase. Examples of well known phase transitions
where the formation of ordered structures are driven solely by an
increase in entropy are the liquid to crystal transition in hard
spheres~\cite{Alder57}, the isotropic to nematic~\cite{Onsager49}
and the nematic to smectic transition in
hard-rods~\cite{Hosino79,Frenkel88a}. Because of their high degree
of monodispersity, and because of the dominant role of steric
repulsion in the pair-potential, colloidal suspensions of
polystyrene latex and rod-like viruses have often been used as
experimental model systems for the study of entropy induced
ordering in hard-sphere~\cite{Forsyth78,Piazza93,Rutgers96} and
hard-rod systems, respectively~\cite{Fraden95,Meyer90}.

A natural extension of the above work is to the phase behavior of
mixtures, with a number of recent experimental and theoretical
studies focusing on the phase behavior of binary mixtures of
hard-spheres~\cite{Biben91,Dinsmore95,Imhof95,Eldrige93,Asakura58,Gast86,Meijer94,Lekkerkerker92,Dijkstra98,Mao95}.
We have recently begun work on less studied systems that closely
approximate hard-rod/hard-sphere and hard-rod/polymer
mixtures~\cite{Warren94,Bolhuis97,Lekkerkerker94,Bruggen98,Lekkerkerker95,Bolhuis94,Vliegenthart99}.
As a model for hard-rods we used either $\it{fd}$ or TMV virus, as
hard-spheres we used polystyrene latex, and as polymers we used
poly(ethylene-oxide) with varying molecular
 weights~\cite{Adams98,Adams98a}. The part of the phase diagram
explored consisted of pure rods in either the isotropic, nematic,
or smectic phase to which a small volume fraction of spheres or
polymers  was added. Remarkably, besides the expected uniform
mixtures and bulk demixing, we also observed a variety of
microphases for a wide range of sphere sizes and
concentrations~\cite{Adams98}. In microphase separation the system
starts separating into liquid-like regions that are rich in either
spheres or rods. However, unlike bulk demixing where rod and sphere
rich regions grow until reaching macroscopic dimensions, in
microphase demixing these liquid-like regions increase only to a
critical size after which they order into well defined three
dimensional equilibrium structures. One of the micro-separated
phases observed, named the lamellar microphase, consists of
alternating two-dimensional liquid-like layers of rods and spheres
and is the subject of theoretical analysis in this paper.

In this paper we use the second virial approximation first studied
by Koda et. al.~\cite{Koda96} to examine the influence of molecular
parameters such as shape and size, on the phase behavior of
rod/sphere mixtures. As the second-virial theory is approximate in
nature, we validate the theoretical predictions by comparing them
with either computer simulations or experimental results. The
remainder of this paper is organized as follows:  In section
\ref{Rods_and_Sphere} we formulate  the second virial approximation
for the rod/sphere mixture. The general features of the phase
diagram are discussed and a physical picture of the factors
responsible for the enhanced stability of the layered phase due to
the presence of spheres is presented. In Section
\ref{Different_Length} the influence of varying the spherocylinder
length on the phase behavior of spherocylinder/sphere mixtures is
studied using computer simulations and the results are compared to
theoretical predictions. Section \ref{Different_Diameter} examines
how changes of the sphere diameter influence the phase behavior of
spherocylinder/sphere mixtures. Finally in Section \ref{Conclusion}
we present our conclusions.

\section{General features of a phase diagram of a spherocylinder-sphere mixture}
\label{Rods_and_Sphere}

Although the equilibrium phases of all hard particle fluids are
determined by maximizing the entropy, ordering transitions are
still possible because the expression for the total entropy, or
equivalently free energy, splits into two parts. The ideal
contribution to the entropy is of the form $\rho \ln \rho$, where
$\rho$ is the density distribution function. This contribution to
the entropy attains a maximum for a uniform density distribution
and therefore always suppresses transitions from uniform to
modulated phases. In contrast, excluded volume entropy sometimes
increases with increasing order and therefore drives the system
towards a modulated phase. In this paper we use a highly simplified
second virial approximation to calculate the excluded volume
entropy.

 The equilibrium phase in a spherocylinder/sphere mixture is
determined by four parameters: length over diameter of a
spherocylinder $(L/D_{sc})$, diameter of spherocylinder over
diameter of sphere ($D_{sc}/D_{sp}$), total volume fraction of
spheres and spherocylinders ($\eta$) and partial volume fraction of
spheres ($\rho_{sp}$). To help us in interpretation of our results
we first define the slope

\begin{equation}
\label{slope}
\tau=\lim_{\rho_{sp} \rightarrow 0} \frac{\eta(\rho_{sp})-\eta(0)}{\rho_{sp}}
\end{equation}

\noindent where $\eta(\rho_{sp})$ is the total volume fraction of the
 rod-sphere mixture at the layering transition after the
introduction of spheres at partial
 volume fraction $\rho_{sp}$. A positive value of $\tau$ implies
that adding a second component stabilizes the nematic phase by
displacing the smectic transition to higher densities. For the case
when both components are spherocylinders of different lengths but
with the same diameter, slope $\tau$ is positive if the ratio of
lengths is less then approximately 7 \cite{Koda94,Stroobants92}. In
the same manner, negative values of $\tau$ imply that the second
component stabilizes the smectic phase. There are predictions of a
negative value of $\tau$ in a bidisperse rod mixture when the ratio
of rod lengths is large enough~\cite{Koda94}, or when added rods
have a larger diameter~\cite{Roij96b}. In this section we focus on
the phase behavior of the spherocyinder-sphere mixture for the
specific microscopic parameters $L/D_{sc}=20$ and $D_{sc}/D_{sp}=
1$. We present a physical picture of excluded volume effects that
are responsible for the enhanced stability of the lamellar phase.
In the next two sections we extend our study on how changes in the
molecular parameters $L/D_{sc}$ and $D_{sc}/D_{sp}$ modify the
phase behavior and in particular, their influence on the magnitude
and sign of the slope $\tau$.

\subsection{Second virial approximation}

\begin{figure}
\centerline{\epsfig{file=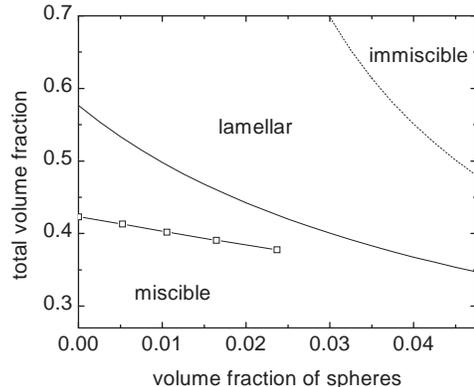,width=6.25cm}}
\caption{\label{PhaseDiagramTheory} Stability boundaries for a
mixture of perfectly aligned spherocylinders $(L/D_{sc}=20)$ and
spheres with equal diameter ($D_{sc}/D_{sp}=1$). The full line
indicates the theoretical prediction of the volume fraction at
which the system becomes unstable towards lamellar fluctuations.
The dashed line indicates instability towards demixing into two
macroscopically distinct phases. Squares are results of computer
simulations at which the layering transition is observed.
Theoretically, the periodicity associated with a one-dimensional
lamellar instability continuously grows and diverges as the system
completely phase separates. Illustrations of the miscible and
lamellar phases are shown in Fig. 5a and 5b, respectively.}
\end{figure}

The second virial approximation for a mixture of perfectly aligned
spherocylinders and spheres of equal diameter was proposed by Koda,
Numajiri and Ikeda~\cite{Koda96} and is generalized for arbitrary
$L/D_{sc}$ and $D_{sc}/D_{sp}$ in the appendix. It was previously
shown that the second virial approximation described qualitatively
the formation and various features of the smectic phase of hard
rods~\cite{Hosino79,Mulder87,Roij95,Schoot96}. Here we study how
the addition of spheres perturbs the formation of the smectic
phase. Since the sphere volume fraction is very low we expect that
the second virial approximation is still qualitatively correct for
these mixtures. We consider a sinusoidal perturbation from the
uniform density for both spherocylinders and spheres. From
equations (\ref{MixFreeEnergyDifference}) and
(\ref{AmplitudeRation}) in the appendix we obtain the free energy
difference between the uniformly mixed  and layered state in a
spherocylinder/sphere mixture:

\begin{equation}
\label{LinearFreeEnergy}
\delta F = a_1^2 \left(S_{11} -2 \frac {a_1}{a_2} S_{12} + \left( \frac {a_1}{a_2} \right) ^{2} S_{22} \right) = 0
\end{equation}.

The phase diagram obtained within this approximation for
microscopic parameters $L/D_{sc}=20$ and $D_{sc}/D_{sp}=1$ is shown
in Fig. \ref{PhaseDiagramTheory}. From the phase diagram we see
that the first prediction of the model is that spheres, upon
addition to a smectic phase, will preferentially occupy space
between smectic layers and therefore create a stable
micro-separated lamellar phase. The second prediction is that the
total volume fraction at which the system undergoes a transition
from a uniform miscible state to a layered lamellar state is
lowered by increasing the partial volume fraction of spheres. This
implies that the slope $\tau$ is negative for this particular
spherocylinder/sphere mixture and we conclude that in this case
{\it spheres enhance the layering transition}.

\begin{figure}
\centerline{\mbox{\epsfig{file=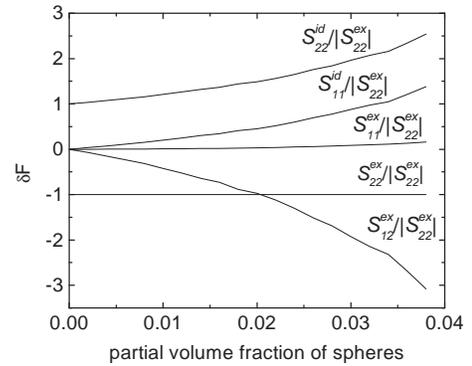,width=6.0cm}}}
\caption{\label{LD10FreeEnergyDiff} Term-by-term dependence of the free
energy difference between the miscible and lamellar phases (Eq.
\ref{LinearFreeEnergy}) as a function of the partial volume
fraction of spheres for $L/D_{sc}=20$ and $D_{sc}/D_{sp}=1$. The
$S_{11}^{id}$ and $S_{22}^{id}$ terms are the sphere and
spherocylinder ideal contributions to the total free energy
difference between the layered and uniform states. $S_{11}^{ex}$,
$S_{12}^{ex}$ and $S_{22}^{ex}$ are excluded volume contributions
to the free energy due to sphere-sphere, spherocylinder-sphere and
spherocylinder-spherocylinder interactions respectively. Since from
our analysis we cannot determine the amplitude in Eq.
\ref{LinearFreeEnergy} we plot amplitude independent ratios of each
of five components of the free energy with different origins to the
spherocylinder-spherocylinder excluded volume interactions. The
stability condition is that $\delta F=0$, so for any value of
partial volume fraction of spheres $\rho_{sp}$ the sum of the five
contributions to $\delta F$ is zero. $\delta F$ of the ideal terms
are positive, hence they stabilize the uniform, miscible nematic
state, while the free volume terms are negative, favoring the
lamellar state. The excluded volume sphere-sphere term
$(S_{11}^{ex})$is negligible and the spherocylinder-sphere
$(S_{12}^{ex})$ term dominates the transition.}
\end{figure}

We can assign a simple physical origin to every term given in Eq.
(\ref{LinearFreeEnergy}) above and Eq.(\ref{matrix}) of the
Appendix. The parts of the spherocylinder-spherocylinder
interaction term $S_{22}$ and sphere-sphere term $S_{11}$ that
scale as $\eta$ are due to the ideal ({\it id}) contribution to the
free energy, also known as the entropy of mixing and are denoted as
$S_{22}^{id}$ and $S_{11}^{id}$, respectively. The terms having a
$\eta^2$ dependence in $S_{22}$, $S_{12}$, $S_{11}$ are due to the
spherocylinder-spherocylinder, spherocylinder-sphere and
sphere-sphere excluded volume ({\it ex}) interaction, respectively
and are denoted as $S_{22}^{ex} $, $S_{12}^{ex}$ and $S_{11}^{ex}$.
Since the instability is defined as $\delta F (\eta_c,k_c)=0$, at a
critical density $\eta_c$ and at a critical wavevector $k_c$ all
individual contributions to the free energy difference in Eq.
(\ref{LinearFreeEnergy}) must add up to zero. In Fig.
\ref{LD10FreeEnergyDiff} we show the value of all terms with
distinct physical origins at the instability density $\eta_c$ and
wavevector $k_c$ as a function of partial volume fraction of
spheres. Since from our analysis we cannot determine the absolute
amplitude of $a_1$  we only plot the ratios of all free energy
components to the absolute value of the
spherocylinder-spherocylinder excluded volume $\mid
S_{22}^{ex}\mid$. If we set the partial volume fraction of spheres
to zero $(\rho_{sp}=0)$ in Eq.(\ref{LinearFreeEnergy}) we obtain an
equation whose solution indicates the nematic-smectic stability
limit in a pure suspension of aligned
spherocylinders~\cite{Mulder87}. For these conditions the only two
nonzero components of free energy are $S_{22}^{ex}$, which is
negative and therefore drives the transition and $S_{22}^{id}$,
which is positive and therefore suppresses the transition. As we
start increasing the partial sphere volume fraction $\rho_{sp}$,
the spherocylinder-sphere free volume term $S_{12}^{ex}$ rapidly
assumes large negative values as evidenced by the rapidly
decreasing ratio of $S_{12}^{ex}/\mid S_{22}^{id}
\mid$. This implies that layering the mixture significantly
decreases the excluded volume that is due to the
spherocylinder-sphere interaction.

\begin{figure}
\centerline{\mbox{\epsfig{file=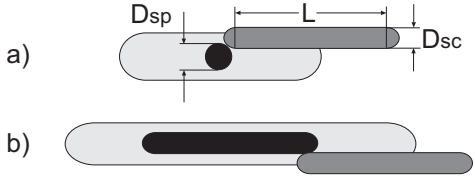,width=6.25cm}}}
\caption{\label{RodSphereExcludedVolume} a) Volume excluded to the center
of mass of a spherocylinder (sc) due to the presence of a sphere
(sp) is indicated by light shading b) Volume excluded to the center
of the mass of a second spherocylinder due to the presence of the
first. Replacing a spherocylinder by a sphere decreases the
excluded volume by approximately a factor of two, but it decreases
the total volume fraction much more since the volume of a
spherocylinder with large $L/D_{sc}$ is greater than the volume of
a sphere with diameter $D_{sc}$. The comparatively large excluded
volume between a sphere and a spherocylinder is the reason for the
enhanced formation of the lamellar phase}
\end{figure}

\begin{figure}
\centerline{\epsfig{file=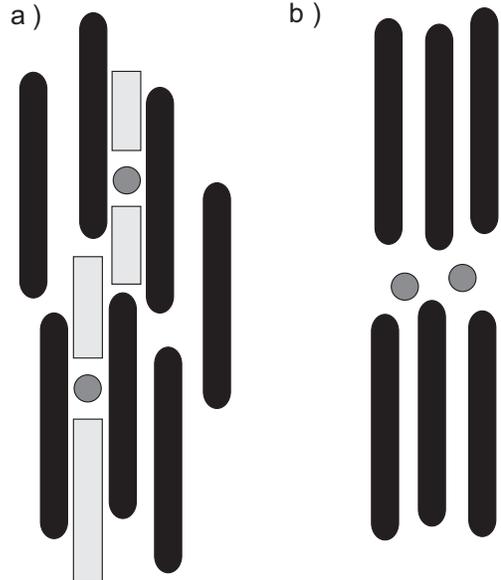,width=6.5cm}}
\caption{\label{RodSphereLamellar} a) A schematic illustration of
excluded volume effects in a nematic phase in a
spherocylinder/sphere mixture. In the nematic or miscible phase
each sphere creates a large excluded volume around it, indicated by
gray areas, that is inaccessible to spherocylinders. b) When the
system undergoes a transition to a layered phase, the large
excluded sphere-spherocylinder volume vanishes since the
probability distribution severely limits the number of ways that
spheres are allowed to approach spherocylinders.  }
\end{figure}

We can use the information gained from the second virial
approximation to obtain a clear physical picture of excluded volume
effects in spherocylinder/sphere mixtures and explain the enhanced
stability of the lamellar phase. Taking any single spherocylinder
in a uniform spherocylinder/sphere mixture and replacing it by two
spheres will leave the value of excluded volume virtually
unchanged. The reason for this lies in the fact that the volume
excluded to the spherocylinder due to the presence of a sphere with
equal diameter, under the constraint of uniform packing, is a
spherocylinder with diameter $2D_{sc}$ and length $(L+2D_{sc})$
where $L$ and $D_{sc}$ are defined in Fig.
\ref{RodSphereExcludedVolume}. However, the excluded volume between
any two spherocylinders with large $L/D_{sc}$ is only about twice
this value as illustrated in Fig. \ref{RodSphereExcludedVolume}.
Although replacing spherocylinders by spheres in such a manner
leaves the excluded volume almost unchanged, it does significantly
decrease the total volume fraction of the mixture since the volume
of two spheres is much smaller then the volume of a spherocylinder
with large $L/D_{sc}$. Therefore in the spherocylinder/sphere
mixture we encounter excluded volume problems similar to those
found in a pure spherocylinder solution, but at a lower total
volume fraction. As in pure spherocylinders, the system reduces the
excluded volume by undergoing a transition to a layered phase. The
excluded volume is reduced in the lamellar state because a periodic
density distribution forces spheres and spherocylinder into
alternate layers thus decreasing the probability of the very
unfavorable sphere-spherocylinder contacts as illustrated in Fig.
\ref{RodSphereLamellar}. This explains the large decrease in the
value of the $S_{12}^{ex}$ term at the lamellar transition that we
observe in the second virial theory. This term is responsible for
the enhanced stability of the lamellar phase in a
sphere/spherocylinder mixture. In conclusion, it is the inability
to efficiently pack a uniform mixture of spherocylinders and
spheres, as reflected in the large spherocylinder/sphere excluded
volume term, that destabilizes the nematic phase and enhances the
formation of a layered phase.

\begin{figure}
\centerline{\epsfig{file=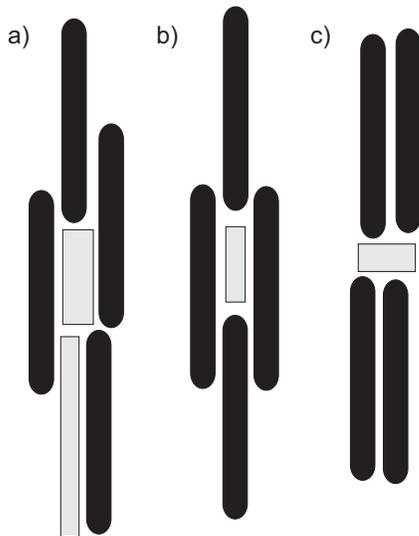,width=5.5cm}}
\caption{\label{NematicSmectic} a) A schematic example of a typical
configuration of spherocylinders in a dense nematic phase. Since
the nematic phase is characterized by a uniform density
distribution this results in inefficient packing and large excluded
volume between spherocylinders both along their radial and axial
directions. This large and unfavorable excluded volume is indicated
by lightly shadowed areas. b) An illustration of a typical
configuration of spherocylinders in a columnar phase where the
excluded volume between spherocylinders is lower compared to the
nematic phase at the same density, and the ideal part of free
energy is higher. In a columnar phase the spherocylinders are
forced into registry as one spherocylinder occupies space right
above or below another one. Therefore the columnar phase is
characterized by two dimensional order in the plane perpendicular
to the spherocylinder's long axis and one dimensional disorder
parallel to the long axis. c) A representative configuration of
spherocylinders in a smectic phase, which is characterized by
one-dimensional order along the spherocylinder's long axis and
two-dimensional disorder in the perpendicular directions. Both
theory and experiment indicate that the columnar phase is always
metastable with respect to the smectic phase
\protect\cite{Mulder87,Veerman91}. }
\end{figure}

 An alternate way to think about the formation of a layered phase
is to focus on the effects of spherocylinder ends~\cite{Wen87}. The
nematic phase in our simplified model is characterized by random
distribution of spherocylinders along their axial and radial
direction as illustrated in Fig. \ref{NematicSmectic}. This end
effect is responsible for the formation of the smectic phase, which
is characterized by a periodic density distribution. In similar
fashion, introducing a sphere into the nematic phase will have the
same effect on the surrounding spherocylinders as another
spherocylinder end. Therefore adding spheres very effectively
increases the density of ``spherocylinder ends" and decreases the
total volume fraction. To resolve the difficulties in efficient
packing due to these extra ``spherocylinder ends", the mixture
layers at a lower total volume fraction.

\subsection{Monte Carlo Simulation}

In the previous section we discussed two predictions of the second
virial theory for a spherocylinder/sphere mixture with
$L/D_{sc}=20$ and $D_{sp}/D_{sc}=1$; the existence of the lamellar
phase and the enhanced stability of the lamellar phase when
compared to a smectic phase of pure spherocylinders. Our results
are in agreement with previous studies of Koda et.
al.~\cite{Koda96}. However, the second virial approximation is
highly approximate and there is reasonable concern about the
influence of higher terms on the topology of the phase diagram. To
support their conclusions Koda et. al. performed computer
simulations, which indicated the existence of an lamellar
phase~\cite{Koda98,Koda96}. Still, the question of whether spheres
simply fill the voids between layers in an already formed smectic
phase, or actually induce layering at lower total volume fraction
was not addressed. In this section, using Monte Carlo simulations
we address the question of the influence of adding spheres on the
phase behavior of spherocylinders by determining the slope $\tau$
in Eq. \ref{slope} in a mixture of spherocylinders and spheres with
parameters $L/D_{sc}=20$ and $D_{sp}/D_{sc}=1$

A Monte Carlo simulation of a mixture of hard-spheres and perfectly
aligned hard-spherocylinders was performed at constant pressure and
number of particles~\cite{Frenkel96}. Most simulations contained
392 spherocylinders and a variable number of spheres. To check for
finite size effects we also ran simulations with 784
spherocylinders, but saw no significant difference in the  results
obtained. In one sweep, pressure was increased
 from a dilute homogeneous mixture up to a well ordered, dense
smectic or lamellar phase. At each value of the pressure, the
density of spheres and spherocylinders and their corresponding
smectic order parameter were measured after the system was allowed
to equilibrate. Identical results were obtained when the  pressure
was slowly decreased from a initially dense phase composed of
alternating layers of spherocylinders and spheres to a dilute
homogeneous mixture.

 Besides lamellar transitions there is a possible demixing
transition where spherocylinders and spheres phase separate into
macroscopically distinct phases. However, once a layered phase is
formed the exchange of spheres between layers drops to a negligible
amount, leaving open the possibility that system would undergo a
demixing transition, but is stuck in a lamellar phase, which is
only a metastable state. To find out the location of the demixing
transition it is necessary to measure the chemical potential of
both spherocylinders and spheres in a spherocylinder/sphere
mixture~\cite{Koda99}. This possibility was not examined in this
work, primarily because we are only interested in how low
concentrations of spheres perturb the formation of the layered
phase. Therefore it is reasonable to expect that at a very low
volume fraction of spheres, the lamellar transition is going to be
more stable than the demixing transitions as predicted by the
second virial theory.

\begin{figure}
\centerline{\mbox{\epsfig{file=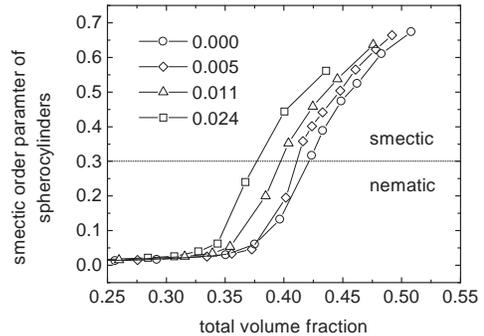,width=6.25cm}}}
\caption{\label{SOPvsdensity} Smectic order parameter obtained from
Monte Carlo simulations is plotted against the total volume
fraction for spherocylinders with $L/D_{sc}=20$. From right to
left, the partial volume fraction of spheres ($\rho_{sp}$)
increases from 0\% to 2.4\% as indicated by the legend. The phase
diagram is reconstructed from this data by defining a phase as
layered when the spherocylinder order parameter reaches a value of
0.3. }
\end{figure}

A plot of the smectic order parameter for spherocylinders with
$L/D_{sc}=20$ as a function of increasing total density for
different partial volume fractions of spheres is shown in Fig.
\ref{SOPvsdensity}.
 As the system approaches a certain critical density we observe a
rapid non-linear increase in the smectic order parameter that we
interpret as a signature of the nematic to smectic phase
transition. This critical density shifts to lower values of the
total volume fraction as the partial volume fraction of spheres is
increased. To reconstruct a phase diagram from the above data we
define a phase as layered when its smectic order parameter reaches
a value of 0.3~\cite{note}. For a pure spherocylinder suspension
this value yields good agreement with previous studies of the
volume fraction of the nematic-smectic phase
transition~\cite{Stroobants87}. Since we are mostly interested in
the qualitative behavior of a spherocylinder/sphere mixture this
method should suffice our purposes. Using this phenomenological
rule, the phase diagram for a mixture of spherocylinders and
spheres ($L/D_{sc}=20$, $D_{sc}/D_{sp}=1$) is reconstructed and
compared to the second virial theory in Fig.
\ref{PhaseDiagramTheory}. An immediate conclusion drawn from Fig.
\ref{PhaseDiagramTheory} is that adding spheres to aligned
spherocylinders enhances the stability of the lamellar phase, which
is indicated by the negative value of slope $\tau$, in agreement
with the prediction of the second virial approximation.

\section{The effects of spherocylinder length on the phase diagram}
\label{Different_Length}

\begin{figure}
\centerline{\mbox{\epsfig{file=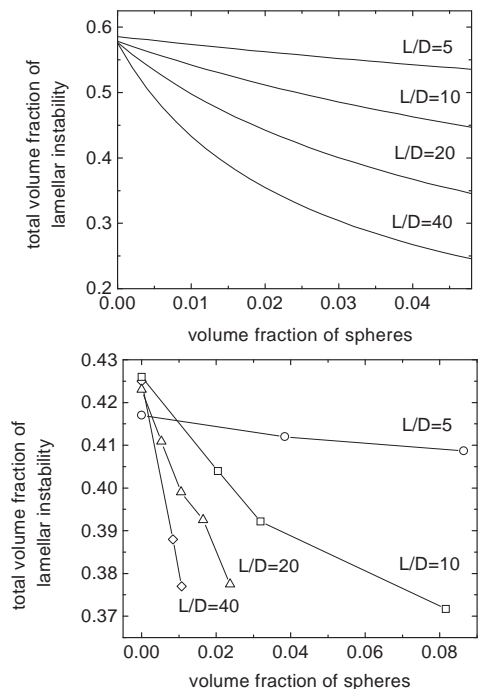,width=6.25cm}}}
\caption{\label{DifferentLD} a) Prediction from the second virial theory
for the total volume fraction $\eta$ of the lamellar instability as
a function of sphere partial volume fraction ($\rho_{sp}$) for
spherocylinders with different $ L/D_{sc}$ ratios. The diameter of
spherocylinders is kept constant and is equal to the diameter of
the spheres. b) Results from Monte Carlo simulations for the
lamellar instability of spherocylinders as a function of partial
volume fraction of spheres for same conditions as in Fig.
\ref{DifferentLD}a. The volume fraction at the phase transition was
defined as having a smectic order parameter of spherocylinders
equal to 0.3}
\end{figure}

Next we proceed to investigate the influence of varying the
spherocylinder length on the magnitude of slope $\tau$. The
predictions of the second virial theory for the nematic-lamellar
instability are shown in Fig. ~\ref{DifferentLD}a. The second
virial theory clearly predicts increasing stability of the lamellar
phase with increasing length of spherocylinder. To verify this
prediction we repeated Monte Carlo simulations for spherocylinders
with different $L/D_{sc}$ and used the same rule as before to
identify the volume fraction of the nematic-lamellar transition.
The simulation results for the location of the nematic to layered
transition are shown in Fig. \ref{DifferentLD}b. We can conclude
that our simulations confirm predictions of the second virial model
and that the length of the spherocylinder is an important parameter
in forming the lamellar phase, with longer spherocylinders showing
an increasing tendency to form a layered phase at a lower volume
fraction of added spheres.

Using the physical picture of the excluded volume effects developed
in the previous section provides a natural explanation for our
simulation results in Fig. \ref{DifferentLD}. With increasing
spherocylinder length the excluded volume due to the
spherocylinder-sphere interaction grows proportionally to the
spherocylinder length and consequently the value of the
$S_{12}^{ex}$ term increases in magnitude. As we have seen before,
the larger the $S_{12}^{ex}$ term, the more likely it is for the
system to form a layered phase.

It is interesting to consider the limit of spherocylinders with
infinite aspect ratio. In the density regime of the nematic-smectic
transition, this model can be mapped onto a system with skewed
cylinders with an aspect ratio close to one. The nematic-smectic
transition in this model has been studied
numerically~\cite{Bolhuis97b,Polson97}. If we consider the addition
of spheres to this system, then the same affine transformation that
maps the infinite spherocylinders onto squat, skewed
spherocylinders, will map the spheres onto infinitely thin,
parallel disks. As the disks are infinitely thin, they do not
interact with each other but only with the cylinders. Inside the
nematic phase, most volume is excluded for these disks. However, in
the smectic phase, there is ample space for the disks between the
layers. In fact, the stronger the layering, the larger the
accessible volume. Hence in this limit, the addition of spheres
will strongly stabilize the smectic phase.

\section{The effects of sphere diameter on the phase diagram}
\label{Different_Diameter}

\begin{figure}
\centerline{\mbox{\epsfig{file=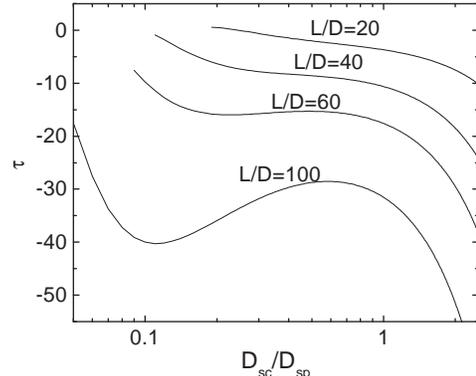,width=6.25cm}}}
\caption{\label{DifferentDR} Theoretical prediction for the
stability criterium of the lamellar phase $\tau$ in Eq. \ref{slope}
as a function of spherocylinder (sc) to sphere (sp) diameter ratio
for four spherocylinders with different $L/D_{sc}$. The negative
value of slope $\tau$ indicates that spheres of that particular
size enhance the layering transition. Larger negative values of
$\tau$ implies the formation of the lamellar phase at a lower total
volume fraction.}
\end{figure}

In this section we investigate the influence of sphere diameter on
the value of slope $\tau$. Fig. \ref{DifferentDR} shows the
prediction of the second virial theory for the dependence of slope
$\tau$ on the ratio of spherocylinder to sphere diameter
$(D_{sc}/D_{sp})$ for spherocylinders with different $L/D_{sc}$. In
section A we examine the phase behavior of sphere/spherocylinder
mixtures when the sphere diameter is smaller then spherocylinder
diameter and in section B we examine the other case when the sphere
diameter is larger then the spherocylinder diameter. In our model
the presence of the spheres cannot alter the orientational
distribution function of spherocylinders, which are always
perfectly parallel to each other. It is reasonable to expect that
this assumption holds for spheres smaller then the spherocylinder
length, but as a sphere becomes larger then the spherocylinder
length, long wavelength elastic effects start to dominate the
behavior of the system and hard spherocylinders will tend to align
parallel to the surface of the sphere~\cite{Poniewierski88}.
Therefore in Fig. \ref{DifferentDR} we plot the values of slope
$\tau$ only for those values of $D_{sc}/D_{sp}$ for which our
assumptions are at least qualitatively correct. As we increase the
sphere size beyond this limit our model describes a highly
artificial system of large spheres and parallel spherocylinders. In
this regime we observe oscillations in the value of slope $\tau$
similar to what is observed in binary mixtures of parallel
spherocylinders~\cite{Koda94}.

\subsection{Sphere diameter smaller than spherocylinder diameter}

In the regime where $D_{sc}/D_{sp}>1$  (for spherocylinders of any
$L/D_{sc}$), decreasing the sphere size increases the stability of
the lamellar phase as indicated by the increasing negative value of
slope $\tau$ seen in the right hand side of Fig. \ref{DifferentDR}.
This prediction of the theory has a simple explanation in our
picture of excluded volume in a sphere/spherocylinder mixture. If
we halve the sphere radius $D_{sp}$, while keeping constant the
volume fraction of spheres, we increase the number of spheres eight
times. At the same time, the result of reducing the sphere size is
to decrease the excluded volume of the spherocylinder-sphere
interaction. However, the eightfold increase in the number of
spherocylinder-sphere interactions more then compensates for the
decrease in excluded volume between the sphere and spherocylinder
and consequently the magnitude of $S_{12}^{ex}$ increases with
decreasing sphere diameter. This leads to the increased  stability
of the layered phase with decreasing sphere size.

It becomes difficult to verify this prediction using computer
simulations. As the sphere size decreases at constant total volume
fraction $\eta$, the number of particles in a simulation rapidly
reaches the order of thousands requiring simulation times that are
prohibitively long. As the ratio of spherocylinder to sphere
diameter $(D_{sc}/D_{sp})$ was varied within the accessible range
between 0.5 to 2 we did not observe any changes in the value of
slope $\tau$ that were larger than our measurement error. Larger
and longer simulations  are needed for a careful analysis of
spherocylinder/sphere mixtures with extreme values of the ratio
$D_{sc}/D_{sp}$.

\subsection{Sphere diameter larger than spherocylinder diameter}

\begin{figure}
\centerline{\mbox{\epsfig{file=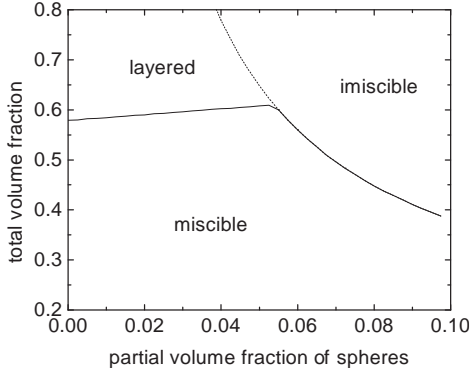,width=6.25cm}}}
\caption{\label{PhaseDiagramLargeSphere} Stability diagram of a
mixture of spherocylinders $(L/D_{sc}=10)$ and large spheres
 $D_{sc}/D_{sp}=0.15$. Unlike  a mixture of small spheres and
spherocylinders (Fig. \ref{PhaseDiagramTheory}), introducing large
spheres displaces the layering transition to higher total volume
fractions indicating a positive value of slope $\tau$. The
structure of the layered phase is also different, with large
spheres fitting in the smectic layer rather than into the smectic
gap. The smectic periodicity associated with the layering
transition
 does not change significantly until the concentration of spheres
is high enough for the system to demix. Then the smectic wavevector
discontinuously jumps to zero.}
\end{figure}

For spherocylinders with  small $L/D_{sc}$, Fig. \ref{DifferentDR}
shows that the magnitude of slope $\tau$ uniformly decreases with
increasing sphere size. Eventually the slope $\tau$ changes sign
and becomes positive, implying that large spheres stabilize the
nematic and not the smectic phase. The phase diagram under
conditions where slope $\tau$ is positive is shown in Fig.
\ref{PhaseDiagramLargeSphere}. The wavevector associated with the
layering transition, indicated with a solid line in Fig.
\ref{PhaseDiagramLargeSphere}, remains at an almost constant value.
Another important point is that the amplitude ratio in Eq.
(\ref{AmplitudeRation}) is positive. This means that the periodic
density modulations of the spherocylinders and spheres are in
phase, which implies that spheres no longer go into the gap between
two spherocylinder layers, but rather fit into the spherocylinder
layer. However, as the partial volume fraction of spheres
$(\rho_{sp})$ is increased further we observe a discontinuous jump
in the wavevector to zero value. This implies that there is a
discontinuous change from a layering to a demixing transition.  As
the demixing transition is reached there is also a change in sign
of the amplitude ratio, which becomes negative and the
spherocylinders and spheres bulk separate. In contrast, the phase
diagram for mixtures of small spheres and spherocylinders shown in
Fig. \ref{PhaseDiagramTheory} looks quite different. The amplitude
ratio for this case is always negative implying formation of the
lamellar phase. Another contrast is that in a mixture of small
spheres and spherocylinders the wavevector associated with the
layering transition decreases in a continuous fashion until it
reaches zero value.

We now examine  the behavior of individual terms in Eq.
(\ref{LinearFreeEnergy}) for a mixture of large spheres and short
spherocylinders shown in Fig. \ref{PhaseDiagramLargeSphere}. Most
notably, we find that at low volume fractions of spheres where the
system undergoes the layering transition, the ratio
$S_{12}^{ex}/S_{22}^{ex}<<1$. This implies that upon layering there
is almost no reduction of the unfavorable sphere/sperocylinder
interaction and that the spherocylinder/spherocylinder interaction
alone drives the formation of the layered phase.  In contrast, for
small spheres this ratio was large and was responsible for enhanced
stability of the lamellar phase as was shown in Fig.
\ref{LD10FreeEnergyDiff}. At a higher volume fraction of large
spheres where the mixture directly bulk phase separates we find
that the ratio $S_{12}^{ex}/S_{22}^{ex}>>1$. This implies, as
expected, that demixing very effectively reduces the unfavorable
sphere/spherocylinder interactions. These results suggest a
physical picture of the excluded volume effect. Unlike small
spheres, large spheres can not fit into the gap between smectic
layers and consequently there is no way to gain free volume by
undergoing the layering transition. As an alternative, to gain free
volume the system bulk phase separates at the lowest volume
fraction of spheres possible.

While for short spherocylinders the magnitude of slope $\tau$
uniformly decreases with increasing sphere size, longer
spherocylinders exhibit a qualitatively different behavior. For a
mixture of spherocylinders with $L/D_{sc}=100$ and spheres with
$D_{sc}/D_{sp}=0.1$
 there is a pronounced  increase in the stability of the lamellar
phase as shown in Fig. \ref{DifferentDR}. By increasing the length
of spherocylinders to even larger values, the region of increased
stability of the lamellar phase shifts to higher values of the
sphere radius. Two conditions emerge, which when satisfied lead to
enhanced stability of the lamellar phase. First, it is necessary
for a sphere to fit between two smectic layers without disturbing
them. This condition is satisfied when $D_{sp}/L \approx 0.1$. The
second condition is that $D_{sp}/D_{sc} >> 1$. It was argued before
that under these condition large spheres are able to induce smectic
correlations amongst neighboring spherocylinders~\cite{Adams98},
which in turn can enhance the formation of the lamellar phase.

 Because of the large size asymmetry it was not feasible to carry
out simulations for mixture of spherocylinders and spheres with
$L/D_{sc}
\approx 100$ and  $D_{sc}/D_{sp} \approx 0.1$. However, these conditions are closely
approximated by recent experiments on rod-like $\it{fd}$
($L=1\mu$m, $L/D_{sc}
\approx$ 100) and polystyrene spheres~\cite{Adams98}. Therefore, we
compare theoretical results of slope $\tau$ for spherocylinders
with $L/D_{sc}=100$ shown in Fig. \ref{DifferentDR} to these
experimental results~\cite{Adams98}. When large spheres $D_{sp}
\approx 1\mu$m, $(D_{sc}/D_{sp} \approx 0.01)$ are mixed with $\it{fd}$ at
any concentration for which the nematic phase is stable, we observe
no formation of the layered phase. Instead, large spheres phase
separate into dense aggregates elongated along the nematic director
indicating that the value of slope $\tau$ is larger then zero. When
the size of the sphere was decreased to $D_{sp}=0.1\mu$m,
$(D_{sc}/D_{sp}\approx 10)$ we observed a transition to a layered
state at a $\it{fd}$ concentration of 20 mg/ml. The formation of a
smectic phase in a pure {\it fd} suspension at the same ionic
strength occurs at 65 mg/ml. The fact that adding spheres
diminishes the rod density by a factor of three indicates a large
negative value of slope $\tau$. As the sphere size was further
decreased $D_{sp}=0.022\mu$m, ($D_{sc}/D_{sp} = 0.46$) there was
again indication of a lamellar phase, but this time at a much
higher concentration of rods of about 50 mg/ml. Thus, although
small spheres still stabilize the layering transition, implying a
negative value of slope $\tau$, the magnitude of slope $\tau$ is
much less for $D_{sc}/D_{sp} \approx 0.46$ then for $D_{sc}/D_{sp}
\approx 0.1$. These qualitative trends of the non-monotonic behavior of slope
$\tau$ with sphere size observed in experiments of {\it
fd}-polystyrene mixtures are very similar to the theoretical
prediction shown in Fig. \ref{DifferentDR} for spherocylinders with
$L/D_{sc}=100$.

\section{Conclusions}
\label{Conclusion}

In this paper we have presented the predictions of the second
virial theory for a mixture of parallel hard-spherocylinders and
hard-spheres undergoing one dimensional microphase separation. We
have been able to verify a number of these predictions using Monte
Carlo simulations. We found that spheres induce layering, which
implies a negative value of the slope $\tau$, which is the change
in total volume fraction of the mixture at the point of
nematic-smectic instability with respect to the partial volume
fraction of added spheres (Eq. \ref{slope})
. At the same time the magnitude of the slope $\tau$ increases
with increasing spherocylinder length. In other words, spheres at
the same partial volume fraction stabilize layering of longer
spherocylinders more then  shorter spherocylinders. Besides this,
the theory predicts an unusual non-monotonic behavior in slope
$\tau$ as a function of sphere to spherocylinder diameter. Although
the physical origin of this effect is not clear, it is intriguing
that similar qualitative trends are observed in experiments of
mixtures of the spherocylinder-like $\it{fd}$ and polystyrene
spheres. However, in real experiments spherocylinders are free to
rotate, are flexible, and have charge associated with them.  Before
quantitative comparisons with experiments are possible it will be
necessary to perform simulations and formulate theories that take
into account these effects mostly ignored in this highly idealized
treatment.

\section{Acknowledgments}

We acknowledge useful discussions with Richard Sear and Bulbul
Chakraborty. We thank Tomonori Koda for critical reading of the
manuscript. This research was supported by NSF DMR-9705336 and NSF
INT-9113312. The work of the FOM Institute is supported by FOM
(``Stichting Fundamenteel Onderzoek der Materie") with financial
aid from NWO (``Nederlandse Organisatie voor Wetenschappelijk
Onderzoek").

\section{Appendix}

A general expression for the free energy of bidisperse mixture at
the second virial level is

\begin{eqnarray}
\label{SecondVirialMixture}
&&\beta F( \rho_1,\rho_2)=\sum_{i=1,2}\int_{V} d{({\bf r})}
\rho_{i}({\bf r}) \ln (\rho_{i} ({\bf r}))- \nonumber \\
&&\frac{1}{2} \sum_{i=1,2}\sum_{j=1,2} \int_{V}d{\bf r_1}
\int_{V}d{\bf r_{2}} \rho_{i}({\bf r_{1}}) \rho_{j}({\bf r_{2}})
f_{i,j}({\bf r_{1},r_{2}})
\end{eqnarray}

where the function $f_{i,j}$ is the overlap function between two
spheres, sphere and spherocylinder or two
spherocylinders~\cite{Koda96}. It attains the value of -1 if two
particles overlap, otherwise it is equal to 0. The terms involving
$\rho \ln \rho$ represent the entropy of mixing while the terms
involving $f_{i,j}$ represent the free volume entropy. Since we are
interested in one dimensional layering we look at the response of
the system to following density pertubation

\begin{eqnarray}
\label{pertubation}
        \delta\rho_{1}(z)=a_{1} \cos(k_{z}z)  \nonumber \\
        \delta\rho_{2}(z)=a_{2} \cos(k_{z}z)
\end{eqnarray}

\noindent The free energy difference between the uniform and perturbed state
is

\begin{equation}
\label{MixFreeEnergyDifference}
\delta F=F(1+\delta \rho_{1}(z),1+\delta \rho_{2}(z))-F(1,1)={\bf \tilde{a}Sa}
\end{equation}

\noindent where ${\bf \tilde{a}}=(a_{1},a_{2})$ and ${\bf S}$ is a two
dimensional stability matrix. To find the limit of stability we
have to solve the equation $\mbox{det}({\bf S})=0$. For latter
convenience we define the following function

\begin{eqnarray}
S(\frac{L}{D_{sc}},\sigma,k) &=&\frac{3 \sin(k \sigma(2+2
\frac{L}{D_{sc}}))}{4k^3}- \nonumber \\
 & &\frac{2 k \sigma \cos(k \sigma(2+2 \frac{L}{D_{sc}})) -\sin(k
 2\sigma \frac{L}{D_{sc}})}{4k^3}
\end{eqnarray}.

\noindent The above expression depends only on geometrical factors and is
related to the Fourier transform of the spherocylinder which is
specified by the excluded volume between a sphere of diameter
$D_{sp}$ and a spherocylinder of length $L$ and diameter $D_{sc}$.
Wavevector $k$ is dimensionless because it is rescaled with the
spherocylinder diameter ($D_{sc}$). The parameter $\sigma$ is
defined as ratio of sphere diameter to spherocylinder diameter
$(\sigma=D_{sp}/D_{sc})$. In the limit of $L/D_{sc}\rightarrow 0$
the above expression reduces to a Fourier transform of a sphere
with unit diameter. The stability matrix ${\bf S}$ for a mixture of
spherocylinders and spheres  has the following form

\vspace{0.5cm}

\begin{eqnarray}
\label{matrix}
&&{\bf S}  =
\left(\begin{array}{cc}
\frac{\displaystyle \eta (1-\rho_{sp})(1+4(1-\rho_{sp}) \eta S(0,1,k))}{\displaystyle 4} \\
\frac{\displaystyle 2\rho_{sp}(1-\rho_{sp})
\eta^2S(\frac{L}{D_{sc}},1+\sigma,k)}{\displaystyle \sigma^6(\frac{2}{3}\frac{L}{D_{sc}}+1)^2}
   \end{array} \right. \nonumber \\
   && \left. \begin{array}{cc} \frac{\displaystyle
   2\rho_{sp}(1-\rho_{sp}) \eta
^2S(\displaystyle \frac{L}{D_{sc}},1+\sigma,k)}{\displaystyle \sigma^6(\frac{2}{3}\frac{L}{D_{sc}}+1)^2}
 \\ \frac{\displaystyle \eta \rho_{sp}(\frac{\displaystyle
 \sigma^6(\frac{3}{2}\frac{L}{D_{sc}}+1)}{4}+ \eta \rho_{sp}
 S(2\frac{L}{D_{sc}},2\sigma,k))}{\displaystyle
 \sigma^6(\frac{2}{3}\frac{L}{D_{sc}}+1)^2}
\end{array} \right)
\end{eqnarray}

\vspace{0.2cm}

\noindent where $\rho_{sp}$ denotes partial volume fraction of spheres and varies
between 0 and 1 while $\eta$ denotes
 total volume fraction. Note that the terms in matrix elements
$S_{11}$ and $S_{22}$  proportional to $\eta$ are due to
configurational entropy while terms proportional to $\eta^2$ are
due to free volume entropy.
 As $k \rightarrow 0$ the condition  $\mbox{det}({\bf S})=0$
reduces to the usual thermodynamic condition for the stability of
the system against bulk phase separation.

To reconstruct the stability diagram from the determinant we slowly
increase the total volume fraction $\eta$. At a certain value of
total volume fraction $(\eta_c)$ the determinant of $\bf S$ will
equal zero for a specific wavevector $(k_c)$. If the wavevector
$k_c$ obtained has a finite value it implies that system is
undergoing a layering transition. On the other hand, the condition
$\mbox{det}({\bf S}) = 0$ when $k_c=0$ implies complete demixing.
Once we obtain values of $\eta_c$ and $k_c$ we can find out the
ratio of amplitudes from the following formula

\begin{equation}
\label{AmplitudeRation}
\frac{a_{1}}{a_{2}}=-\frac{S_{12}(\eta_c,k_c)}{S_{11}(\eta_c,k_c)}.
\end{equation}

\noindent A positive value of the amplitude ratio implies that the
spheres and spherocylinders are in the same layer (the periodic
modulations are in phase), while a negative value implies that the
spheres and spherocylinders intercalate (the periodic modulations
are out of phase).


\begin{thebibliography}{99}
\bibitem{note}
The smectic order parameter is defined as $ \mid \sum_{j} e^{i k_z r_j}
  \mid^2/N^2$ evaluated at $k_z$ for which the expression attains it's maximum
  value, where $r_j$ is the center of mass of the spherocylinder or sphere, and
  $k_z$ is the wavevector along the spherocylinder's long
  axis~\protect\cite{Veerman91}.
\end{thebibliography}
\end{document}